\documentclass[preprint2]{aastex}

\shorttitle{Fragile Binaries Candidates in the SDSS DR8 spectroscopic archive}
\shortauthors{Zhao et al.}

\begin{document}


\title{Fragile Binary Candidates in the SDSS DR8 spectroscopic archive}


\author{J. K. Zhao\altaffilmark{1,2}, T. D. Oswalt\altaffilmark{1},  G. Zhao\altaffilmark{2}}
\email{jzhao@fit.edu}
\email{toswalt@fit.edu}
\email{gzhao@bao.ac.cn}


\altaffiltext{1}{Florida Institute of Technology, Melbourne, USA, 32901}
\altaffiltext{2}{Key Laboratory of Optical Astronomy, National Astronomical Observatories, Chinese Academy of Sciences, Beijing, 100012, China}


\begin{abstract}
We present a catalog of 80 very wide fragile binary candidates (projected separations $>$ 10000 AU) from the SDSS DR8 spectral archive. The pairs were selected based on proper motion, radial velocity, metallicity and photometric parallax criteria. The angular separations of these pairs range from 3$\arcmin\arcmin$ to 250$\arcmin\arcmin$. The peak in the metallicity distribution of these pairs is about -0.5 dex of solar metallicity. Space motions and reduced proper motion diagrams indicate all these pairs are members of the disk. The chromospheric activity index $S$$\rm_{HK}$
of each component in 38 binary candidates having spectra of high signal-to-noise ratio and member stars of three open clusters (NGC2420, M67 and NGC6791) were measured. The $S$$\rm_{HK}$ vs. color relation for these binary
 candidates is consistent with the trend seen in these open clusters. The ages implied by this relation suggest that fragile wide pairs can survive longer than 8 Gyr.
\end{abstract}



\section{Introduction}
Wide fragile binaries by definition have large semimajor axes (a  $\succeq$ 100 AU). Thus, each component may be assumed to have evolved
independently, unaffected by mass exchange or tidal coupling
that complicate the evolution of closer pairs (Greenstein 1986).
It may also be assumed that members of such binaries are coeval.
Essentially, each may be regarded as an open cluster with only
two components. Fragile binaries are important probes of the nature of halo dark matter, the evolution
of the stellar halo, and the metallicities, masses, and ages of field stars (see Chanam\'{e} 2007). To better understand the formation (Kouwenhoven et al. 2010)  and evolution (Jiang $\&$ Tremaine 2010) of wide binaries, large samples are needed.

At present, candidate fragile binaries have been selected mainly
by searching for common proper motion (CPM) pairs. Luyten (1979,
1988) pioneered this technique using Schmidt
telescope plates and a blink microscope. He detected more
than 6000 wide pairs with $|\mu|$ $>$ 100 mas yr$^{-1}$. This method has since been used to find
fragile binaries in the AGK 3 stars by Halbwachs (1986),
in the revised New Luyten Two-tenths catalog (rNLTT; Salim $\&$
Gould 2003) by Chanam\'{e} $\&$ Gould (2004), and
among the Hipparcos stars in the Lepine-Shara Proper
Motion-North catalog (LSPM-N; L\'{e}pine $\&$ Shara 2005; L\'{e}pine $\&$ Bongiorno 2007). All of these studies used
magnitude-limited high proper motion catalogs and thus are limited mostly to nearby stars.

Recent large-scale surveys such as the Sloan Digital Sky
Survey (SDSS; York et al. 2000), the Two Micron All Sky
Survey (2MASS; Cutri et al. 2003), and the UKIRT Infrared
Deep Sky Survey (UKIDSS; Lawrence et al. 2007), have yielded
samples with good photometric data, that are useful in selecting  more distant fragile binaries
when combined with proper motion information. Sesar et al. (2008) searched
the SDSS Data Release Six (DR6; Adelman-McCarthy et al.
2008) for fragile binaries with angular separations up to 30$\arcmin\arcmin$ using a novel statistical technique that minimizes the difference
between the distance moduli obtained from photometric parallax
relations for candidate pairs. They matched proper motion
components to within 5 mas yr$^{-1}$ with absolute proper motions of 15 - 400 mas/yr. Their search identified $\sim$  14,000 total candidates with excellent completeness. However, one third
of them are expected to be false positives. They found pairs in all mass ranges separated by 2000 - 47,000 AU at distances up to 4 kpc. Quinn $\&$ Smith (2009) searched for new wide halo binary stars in the SDSS Stripe 82 that satisfy common proper motion and photometric distance constraints. The projected separations of their pairs range from 0.007 to 0.25 pc. Longhitano $\&$ Binggeli (2010) used an ``angular two point
correlation function" to do a purely statistical study of fragile
binaries in a $\sim$675 deg$^{2}$ field centered at the North Galactic Pole
using the DR6 stellar catalog. Their work predicted that there are more
than 800 binaries with physical separations larger than 0.1 pc but
smaller than 0.8 pc in this field. Dhital et al. (2010) presented a catalog of 1342
very wide (projected separation $\succeq$ 500 AU), low-mass (at least one mid-K to mid-M dwarf component) fragile pairs identified from astrometry, photometry, and proper motions in the SDSS Data Release Seven (DR7; Abazajian et al. 2009). In their catalog 98.35\% were expected to be physical pairs.

These previous fragile binary searches did not use spectral information such as radial velocity (RV) and metallicity. The SDSS provides medium resolution spectra for about one million stars.
We searched for fragile binary candidates using proper motion, RV and metallicity information in the SDSS spectral archive catalog. RV and
metallicity help to eliminate most random optical  pairs.

%

Section 2 presents a discussion of our data selection method. The fragile binary candidates found are discussed in section 3.
Section 4 examines the chromospheric activity (CA) of the candidate pairs found and, for comparison, among SDSS stars in three open clusters. We conclude with a discussion of our findings in section 5.
\section{Sample Selection}


\subsection{Overview of the SDSS Spectroscopic Data}
The SDSS provides homogeneous and deep
(r $<$ 22.5) photometry in five band passes (u, g, r, i, and z; Gunn
et al. 1998, 2006; Hogg et al. 2001; Smith et al. 2002; Tucker
et al. 2006) accurate to $\pm$0.02 mag (rms scatter) for unresolved
sources not limited by photon statistics (Scranton et al. 2002). This sample has a zero-point uncertainty of $\pm$0.02 mag
(Ivezi\'{c} et al. 2004). The SDSS also provides more than half a million stellar spectra with wavelength ranging from 3800 - 9000 $\rm \AA$. RV and metallicity are
 provided in the Table sppParams (Lee et al. 2008). Moreover,  in DR8 (Aihara et al. 2011a), all the stellar spectra obtained with the SDSS spectrograph were reprocessed through an improved stellar parameter pipeline, which improved the accuracy of metallicity estimates for stars up to solar metallicity.
SDSS spectroscopy was carried out
by twin fiber-fed spectrographs collecting 640 simultaneous observations.
Typical exposure times were $\sim$15 - 20 minutes, but
exposures were subsequently co-added for total exposure times of
$\sim$45 minutes, producing medium resolution spectra with R $\sim$
2000 (York et al. 2000). SDSS spectroscopic plates each contained
16 spectrophotometric standard stars, which were selected
by color to be F subdwarf stars. The SDSS spectroscopic
fluxes were calibrated by comparing these standard stars to a grid
of theoretical spectra from  model atmospheres (Kurucz
1993) and solving for a spectrophotometric solution for
each plate.

\subsection{Initial Data Selection}
Our initial sample was selected mainly from three tables in DR8: \textit{specphotoall}, \textit{sppParams} and \textit{propermotions}\footnote[1]{http://www.sdss3.org/dr8/}. The photometry and extinction values are provided in the Table \textit{specphotoall}. The Table \textit{sppParams} presented RV, $T$$\rm_{eff}$, log $g$ and  [Fe/H], while the Table \textit{propermotions} provided the proper motion of each star as matched with the USNO-B survey (Munn et al.
2004). The original data can be accessed through CasJobs\footnote[2]{http://skyservice.pha.jhu.edu/casjobs/}. With the condition flag = `nnnnn' in Table \textit{sppParams} and class = `star'
 in Table \textit{specphotoall}, we obtained a first-cut sample containing 341,528 stars. Some stars with inaccurate photometry, extinction values  and illegal  metallicity value (-99999) were deleted, leaving 303,587 stars.

 Aihara et al. (2011b) described some unexpected errors in the SDSS DR8 data that might cause a systematic shift in proper motion. To test the effect on our fragile binary candidates, we compared our sample's proper motions in the DR7 and DR8 (see Fig. 1). In right ascension we found a systematic shift of 0.086 mas/yr with a scatter of about 3.4 mas/year. In declination there is a systematic shift of 0.096 mas/yr with a scatter of 2.8 mas/yr.

Our DR8 sample included 303,587 stars, but proper motions of only 219,844 of these stars can be found in DR7. We searched for fragile binary candidates among these 219,844 stars using the DR7 proper motions and found 53 candidates using our six constraints (see Sec. 3). Of these, 51 candidates are a subset of those found in the 303,587 star sample from DR8. Thus, the choice of DR7 or DR8 resulted in essentially the same candidate pairs common to both data sets. In order to start with a larger sample we chose to use the 303,587 stars having DR8 proper motion data.



\section{Fragile Binary Candidates Catalog }
From the initial sample, we searched for fragile binaries using the following constraints:

1. Angular separation of 3$\arcmin\arcmin$ $<$ $\theta$ $<$ 250$\arcmin\arcmin$ between two nearby point
sources A and B on the sky were selected, where $\theta$ was calculated from the small
angle approximation:

\begin{eqnarray}
  \theta\simeq\sqrt{(\alpha_{A}-\alpha_{B})^2\cos\alpha_{A}\cos\alpha_{B}+(\delta_{A}-\delta_{B})^2}
\end{eqnarray}

Sesar et al. (2008) constructed two independent samples of candidate fragile binaries: (1)  3$\arcmin\arcmin$ $<$ $\theta$ $<$ 16$\arcmin\arcmin$ and
(2) 5$\arcmin\arcmin$ $<$ $\theta$ $<$ 30$\arcmin\arcmin$.  Dhital et al. (2010) provided a fragile binary catalog with 7$\arcmin\arcmin$ $<$ $\theta$ $<$ 180$\arcmin\arcmin$.
Although fragile binaries have been found at much larger angular separations (up to 900$\arcmin\arcmin$ in Chanam\'{e} $\&$ Gould 2004; 1500$\arcmin\arcmin$
in L\'{e}pine $\&$ Bongiorno 2007; 570$\arcmin\arcmin$ in Faherty et al. 2010), here we limited our maximum angular separation to 250$\arcmin\arcmin$ since the
number of random pairs with larger angular separations becomes unacceptably high in the deep SDSS survey. After this step, 68,414 pairs remained in our sample.

2. The maximum acceptable difference in proper motion, $\Delta|\mu|$ $<$ 6 mas yr$^{-1}$, was adopted where

\begin{eqnarray}
\Delta|\mu|\equiv\sqrt{(\mu_{lA}-\mu_{lB})^2 + (\mu_{bA}-\mu_{bB})^2}
\end{eqnarray}

The proper motions were queried from SDSS database in the Table ProperMotions which was derived by SDSS/USNO-B cross matching (Munn et al. 2004). We adopted the
proper motions from the DR8 catalog which uses SDSS galaxies to recalibrate the USNO-B positions and SDSS stellar astrometry as an additional epoch for improved
proper motion measurements. The typical 1$\sigma$ error is $\pm$3-4 mas yr$^{-1}$ for each star. This is the reason we eliminated pairs with proper motions difference larger than 6 mas yr$^{-1}$. After this step, 22,964 pairs left.

3. The constraint on RV we adopted for selection of a candidate pair was:
    $|\Delta$RV$|$ $<$ 20 km s$^{-1}$.

The RV values are from the Table sppParams, which were measured by cross correlation with the ELODIE (Moultaka et al. 2004)  stellar library. The typical error in RV is smaller than 10 km s$^{-1}$. Hence, we chose $\Delta$RV smaller than 20 km s$^{-1}$. However, for higher RV stars, we only eliminated the pairs with $\Delta$RV larger than 40 km s$^{-1}$. After this step, 6592 pairs remained.

4. We adopted an additional selection constraint based on metallicity, i.e.

$|\Delta$[Fe/H]$|$ $<$ 0.3

The metallicities are also from Table sppParams in the DR8 catalog. Several methods exist to estimate [Fe/H] (Lee et al. 2008). The typical error in [Fe/H] is no more than
$\pm$0.15 dex. Thus, pairs with $|\Delta$[Fe/H]$|$ $>$ 0.3 were regarded as optical pairs. After this step, 3900 pairs remained.

5. We applied an additional candidate selection condition based on photometric distance, i. e.

$\delta$d $<$ 40\%

Physical pairs should have the same distances within the catalog uncertainties. Photometric parallax relations in the literature differ in the methodology used, photometric systems, and the
absolute magnitude and metallicity range for which they are applicable. Not all of them are mutually consistent. Most exhibit intrinsic
scatter of order $\pm$0.5 magnitude or more. We adopted the relation from Ivezi\'{c} et al. (2008), which gives the absolute magnitude
in the r band, M$\rm_{r}$ as a function of color, g-i and [Fe/H], as follows:
\begin{eqnarray}
\rm M_{r}=\rm -5.06+14.32(g-i)-12.97(g-i)^2+  \nonumber \\
\rm 6.127(g-i)^3-1.267(g-i)^4+0.0967(g-i)^5 \nonumber \\
\rm +4.5-1.1[Fe/H]-0.18[Fe/H]^2
\end{eqnarray}
Since the typical uncertainty of photometric distance is more than 20\%, we limited the selection of physical pair candidates to those with a computed distance difference smaller than 40\%. After this step, 2260 pairs remained.

6. A selection criterion based on projected separation (a) was adopted, i. e.,
a $<$ 0.5 pc. Pairs wider than this are believed to dissolve within the age of the Galaxy due to cumulative encounters with giant molecular clouds, distant encounters with other stars, and the Galaxy's tidal
field (Weinberg, Shapiro $\&$ Wasserman 1987). After this step, 80 candidate pairs remained in our final selected sample.

Table 1 lists the physical properties of these 80 pairs. Columns 1-2 list the $\alpha$ and $\delta$; columns 3-6 list the
proper motions; RV and [Fe/H] are given in columns 7-10; Columns 11-16 list the r magnitude, g-i color and spectral type, respectively. The last two columns list the  angular
separations in arcsec and projected separations in pc.

 Fig. 2 presents the angular separation ($\theta$) distribution. It is bimodal with two peaks: $\theta$ = 25$\arcmin\arcmin$ and $\theta$ = 80$\arcmin\arcmin$. We made a statistical analysis of g-r, [Fe/H] and the dispersion of the W space motion ($\sigma$$\rm_{W}$) for the primaries of the fragile pairs in each
peak. The average $<$g-r$>$ of these two peaks are 0.48 and 0.61; the average $<$[Fe/H]$>$ are -0.45 and -0.50; $\sigma$$\rm_{W}$ are 25 km s$^{-1}$ and 22 km s$^{-1}$. Thus, there
are no significant differences in metallicity and $\sigma$$\rm_{W}$ for these two peaks. Only the average $<$g-r$>$ are a little different. Most primaries of fragile pairs in the first peak are G stars, while
most primaries in the second peak are K stars. Although our maximum
angular separation limit was 250$\arcmin\arcmin$, no pair was found with angular separation larger than 190$\arcmin\arcmin$.

 Fig. 3 is the proper motion distribution. Since we did not set a low cut-off for proper motion, nearly 90\% of our candidate pairs have proper motions lower than 13 mas yr$^{-1}$.

 Fig. 4 is the reduced proper motion (RPM) diagram of the pairs, i.e., H$\rm_{r}$ vs. (g-r)$_{0}$, where H$\rm_{r}$ = r$_{0}$+5log$|\rm\mu|$+5. Equivalently, H$\rm_{r}$ = M$\rm_{r}$ +5*log(V$\rm_{t}$)-3.25, where
M$\rm_{r}$ is the absolute magnitude in the r band and V$\rm_{t}$ is the heliocentric tangential velocity in km s$^{-1}$ given by
V$\rm_{t}$ = 4.74$\rm\mu$d. The dotted line indicates the division between the halo and disk, which was set by adopting V$\rm_{t}$ = 220 km s$^{-1}$ and [Fe/H] = -1.5 and the photometric distance given by Ivezi\'{c} et al. (2008). It is clear that our pairs are all disk stars.

 Fig. 5 is the distance distribution of our pairs. All distances are larger than 100 pc. The peak at about 0.85 kpc indicates that our pairs are members of the disk. The
thick solid line is an exponential fit:

 \begin{eqnarray}
\rm N = \rm 17.09*\exp(\frac{-d}{0.81})-0.18
\end{eqnarray}

Note that the scale height implied by this fit (0.81 kpc) is very similar to the generally accepted scale height of the Galaxy's thick disk (0.75 kpc; de Jong et al. 2010).

 Fig. 6 is the metallicity [Fe/H] distribution of our pairs. A peak is evident at [Fe/H]
 $\sim$ -0.5 dex, which provides more evidence that these pairs are disk stars.


 Fig. 7 is a plot of log$\Sigma$N vs. 3logd, which tests the completeness of our sample. $\Sigma$N is the cumulative number of candidates pairs out to a distance d. The straight line corresponds to $\Sigma$N $\sim$ d$^{3}$. As can be seen, the completeness is high only for pairs within about 1160 pc and falls
off abruptly after that. This is
primarily due to the projected separation limit set in our fragile binary search.

 To investigate how the use of the spectroscopic sample in the SDSS influences their identification as possible wide binaries, we randomly selected 160 stars from the SDSS photometric sample. These 160 stars have no spectroscopic observations. Fig. 8 is a comparison between our sample which consists of 160 stars from final 80 candidate pairs having spectra and the random photometric sample of 160 stars. The two samples have almost the same completeness in distance smaller than 1 kpc. At distances larger than 1 kpc, the photometric sample has better completeness than the specscopic sample, presumably because spectra are difficult to obtain in faint stars.

With the proper motion and photometric distance data,
the rectangular velocity components relative to the Sun
for these pairs were then computed and transformed
into Galactic velocity components U, V, and W,
and corrected for the peculiar solar motion
(U, V, W) = (-9, +12, +7) km s$^{-1}$ (Wielen 1982). The UVW-velocity
components are defined as a right-handed system with U positive
in the direction radially outward from the Galactic center,
V positive in the direction of Galactic rotation, and W
positive perpendicular to the plane of the Galaxy in the
direction of the north Galactic pole. The uncertainties of the U, V and W components were calculated based on the estimated errors in proper motion, distance and RV using the equation 2 of Johnson $\&$ Soderblom (1987). Columns 2-7 in Table 2 list the U, V, W velocity components and their uncertainties for each component in binary candidates. The U, V, W differences between two components of each pair are given in columns 10-12.

 The top panel of Fig. 9 shows the U, V velocity contours, centered at
(U, V) = (0, -220) km s$^{-1}$, that represent 1$\sigma$ and 2$\sigma$ velocity ellipsoids
for stars in the Galactic stellar halo as defined by Chiba $\&$
Beers (2000).
 The bottom panel of Fig. 9 shows the Toomre diagram of candidate pairs (Venn et al. 2004). Stars with V$_{total}$ $>$ 180 km s$^{-1}$ are possible halo members. These plots indicate that all our pairs are disk stars.

\section{Chromospheric Activity Measurements of Fragile Binary Candidates}
\subsection{$S$$\rm_{HK}$ Measurement }
For decades chromospheric activity (CA) has been known to inversely correlate with stellar age (Skumanich 1972). Early work by Wilson (1963; 1968) and Vaughan $\&$ Preston (1980) established CaII H$\&$K emission as a useful marker of CA in stars. Following Hall et al. (2007), for each star we computed the flux ratio $S$$\rm_{HK}$:
\begin{eqnarray}
S\rm_{HK}&\equiv&\rm{\alpha\frac{H+K}{R+V}}
\end{eqnarray}
where H and K are the fluxes measured in 2 $\rm{\AA}$ rectangular windows centered on the line cores of CaII H$\&$K; R and V are the fluxes measured in 20 $\rm{\AA}$ rectangular `pseudocontinuum' windows on either side. These
bands are essentially identical to those used in the Mount Wilson chromospheric activity survey program (Baliunas et al. 1995), except
that the bands centered on Ca II H$\&$K are wider (2 $\rm \AA$) than those used at Mount Wilson (1 $\rm \AA$ ) because of the resolution of
SDSS spectra R $\sim$ 2000. Here $\alpha$ is 10, representing the fact that the psendocontinuum windows are 10 times wider than the H$\&$K windows in
wavelength coverage (Zhao et al. 2011).


Since there are 640 fibers in the SDSS spectrograph, there could be some systematic differences among spectra taken in different fibers. Stars with repeated observations provided the opportunity to
 measure the internal consistency of CA measurements. Eight stars with two or more spectroscopic observations were found. The mean $S$$\rm_{HK}$ difference between
spectra for the same objects taken in different fibers is only about $\pm$0.002, thus we conclude the fiber effect can be ignored in our CA analysis.

\subsection{$S$$\rm_{HK}$ Measurements among MS members of Open Clusters}
In order to estimate the ages of fragile binary candidates, three open clusters NGC2420, M67 and NGC6791 were selected for measurements of the $S$$\rm_{HK}$ from the SDSS DR8.
The member stars of the three open clusters  were selected based on the criteria from Smolinski et al. (2011). The age of NGC2420 is about
2.0 Gyr (Von Hippel $\&$ Gilmore 2000). The age of M67 is about 4.05 Gyr (Jorgensen $\&$ Lindegren 2005). The age of NGC6791 is about 8 Gyr (Grundahl et al. 2008).

 Fig. 10 shows the color-magnitude diagrams (CMD) for these three clusters obtained from DR8 data.  The top panel is the CMD for NGC2420, which has 138 dwarf stars. The CMD of this cluster is well-defined. Unfortunately,
most spectra of this cluster have S/N $<$ 40. The middle panel is the CMD for M67. It includes 72 member stars, all of which are dwarf stars whose spectra have signal to noise (S/N) $>$ 50. The bottom panel shows the CMD for NGC6791, which is a very old open cluster. Forty-five member stars with good S/N were found in this cluster. Only two spectra have S/N $<$ 40 {\bf (points indicated by squares in the bottom panel of Fig. 10)}. In this cluster, high mass members have evolved off the main sequence (points indicated by plus sign in the bottom panel of Fig. 10); these were omitted from the following analysis.

Fig. 11 is the $S$$\rm_{HK}$ vs. (g-r)$_{0}$ diagram for these three open cluster member stars. Plus signs represent the member stars of M67; squares are member stars of NGC2420; triangles represent the member stars of NGC6791. The dotted lines are the least-squares fitted lines  for NGC2420 with $\pm$1$\sigma$; the dashed lines are the fitting for M67, while the dashed dot lines are the fitting for NGC6791. All the fitted lines are parabolas. The scatter in NGC2420 is large because the S/N of the spectra  is very
low. The NGC6791 fitted line is concave down only because this evolved cluster has no blue stars on the upper main sequence to define the curvature.

\subsection{The age estimation of fragile binary candidates}
Although the scatter of  $S$$\rm_{HK}$ is appreciable within each cluster, (especially in NGC2420 and NGC6791), the mean relations in Fig. 11 clearly show that $S$$\rm_{HK}$ declines with age. NGC2420 has
stronger CA at each (g-r)$_{0}$ color than the other two clusters, indicating it is the youngest cluster.

 Fig. 12 displays $S$$\rm_{HK}$ vs. (g-r)$_{0}$ for the 38 fragile binaries with S/N $>$ 40. In this figure, the mean relation for the three clusters in Fig. 11 is overplotted. The fitting lines of these three clusters
can be used to roughly gauges the ages of our fragile pairs. Evidently, the age of the lowest two pairs (J204212.9+560534 $\&$ J204228.2+560539; J082555.4+384633 $\&$ J082602.0+384723) are about the same as NGC6791, i.e. 8 Gyr. The semimajor axes of these two pairs are 0.4 pc and 0.2 pc. Most pairs have about
  the same age as M67. Nearly all the pairs' components have consistent CA level for their (g-r)$_{0}$ color, indicating they are indeed coeval. Only one pair, SDSS J213206.3+750645  $\&$ SDSS J213148.24+750552.63 (connected by a thick solid line in Fig. 12), appears to be nonphysical.  Column 15 in Table 2 presents the age consistency of these 38 pairs. `Y' indicates that two components have consistent ages; `N' indicates that inconsistent ages. `NULL' indicates stars that are not part of the sample of these 38 candidates.

A long dash line is the least squares fitting for all the 80 candidate pairs. It lies mostly between 4 Gyr and 8 Gyr. if we suppose there is a linear relation between age and $S$$\rm_{HK}$ at each (g-r)$_{0}$, the mean age of these pairs is about 5 Gyr, i. e., about the same as the sun.

 For each binary candidate, we adopted a rough confidence level for being a physical pair. We set 5 levels: `A', `B', `C', `D' and `E'. The criteria for level `A', `B', `C' and `D' are shown in Table 3. Candidate pairs that do not satisfy `A' to `D' criteria are given level `E'.  Level `A' means that candidate is very likely to be a physical pair, while `E' corresponds to the lowest probability.

\section{Conclusions}
We found 80 fragile binary candidates with low proper motion based on common proper motion, RV, metallicity and photometric distance. All these pairs have very large projected separations. They are all disk stars based on our analysis of their space motions, metallicities and RPM. The S/N of the spectra for half of these pairs are high enough to measure the CA index $S$$\rm_{HK}$.  Measurements of $S$$\rm_{HK}$ for stars in three open clusters allowed us to make a very preliminary estimate of the age of these pairs.
The mean age of these fragile candidate pairs is about 5 Gyr. Our results suggest that at least some fragile pairs (a $\sim$ 0.4 pc) can survive 8 Gyr in the Galactic disk.  Additional more accurate observations are needed to confirm the truly physical pairs among these candidates.

%




\acknowledgments
 We are grateful for constructive comments by the referee that substantially improved our paper. T.D.O. acknowledges support from NSF grant AST-0807919 to Florida Institute of Technology. J.K.Z. and G.Z. acknowledge support from NSFC grant No. 10821061 and 11078019.

Funding for SDSS-III has been provided by the Alfred P. Sloan Foundation, the Participating Institutions, the National Science Foundation, and the U.S. Department of Energy Office of Science. The SDSS-III web site is http://www.sdss3.org/.

SDSS-III is managed by the Astrophysical Research Consortium for the Participating Institutions of the SDSS-III Collaboration including the University of Arizona, the Brazilian Participation Group, Brookhaven National Laboratory, University of Cambridge, University of Florida, the French Participation Group, the German Participation Group, the Instituto de Astrofisica de Canarias, the Michigan State/Notre Dame/JINA Participation Group, Johns Hopkins University, Lawrence Berkeley National Laboratory, Max Planck Institute for Astrophysics, New Mexico State University, New York University, Ohio State University, Pennsylvania State University, University of Portsmouth, Princeton University, the Spanish Participation Group, University of Tokyo, University of Utah, Vanderbilt University, University of Virginia, University of Washington, and Yale University.

\clearpage




\begin{figure}
\epsscale{1.0}
\plotone{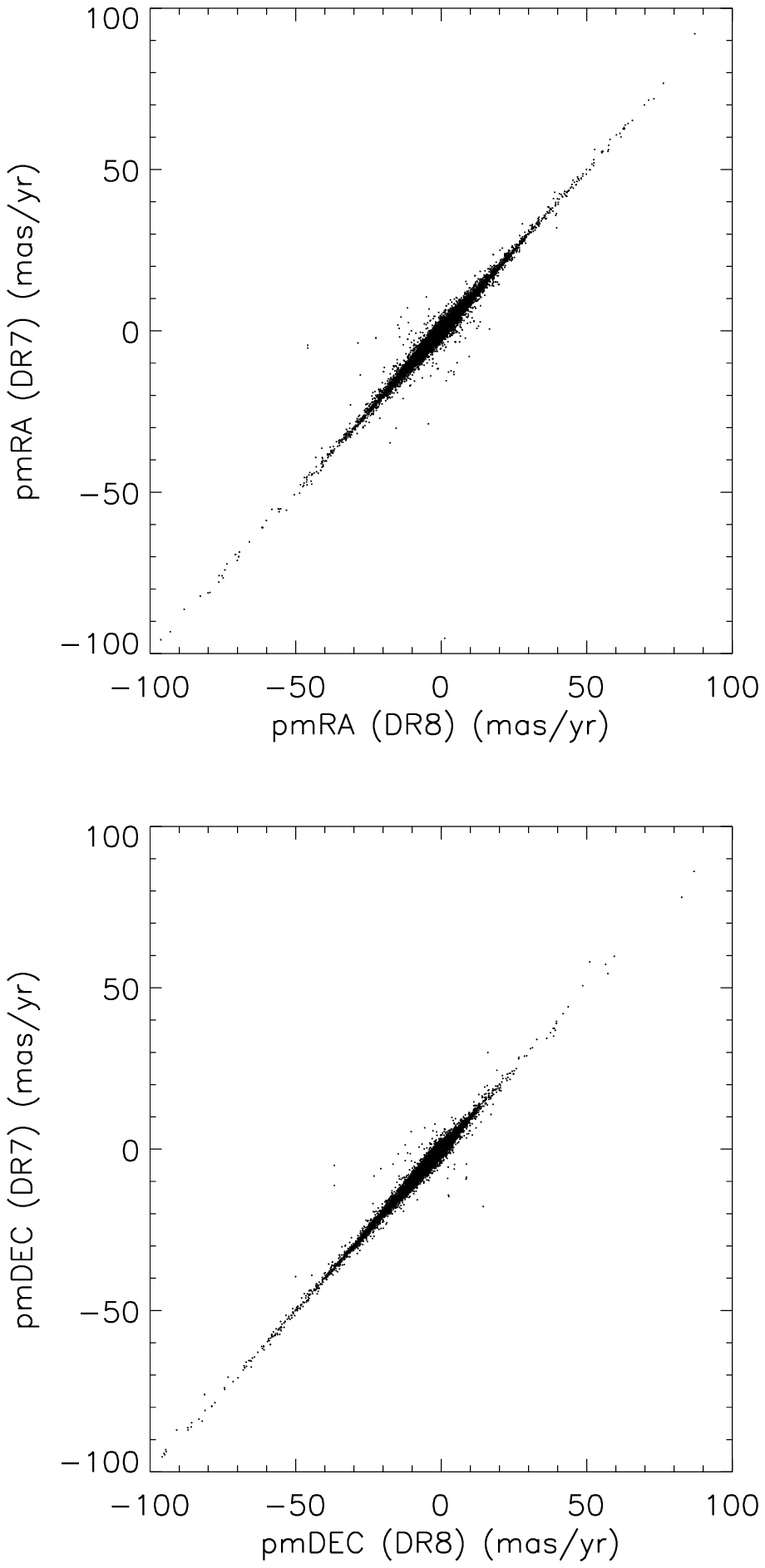}
\caption{Comparison of the DR7 and DR8 proper motions. See the text for an explanation.}
\end{figure}

\begin{figure}
\epsscale{1.0}
\plotone{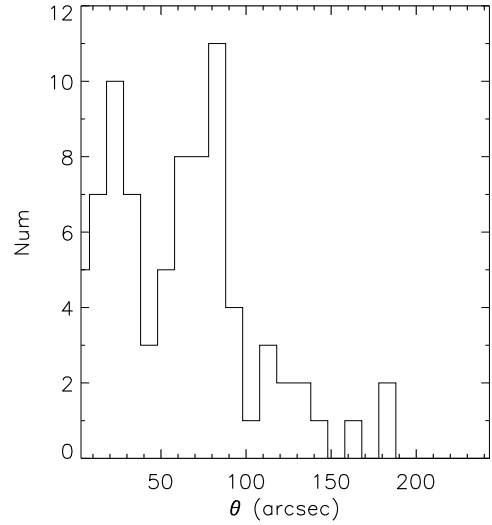}
\caption{Angular separation $\theta$ distribution of our
fragile binaries.}
\end{figure}

\begin{figure}
\epsscale{1.0}
\plotone{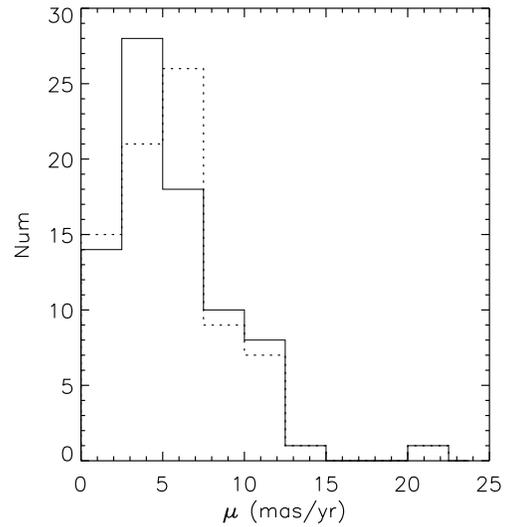}
\caption{Proper motion distribution of these fragile binary candidates. The solid line represents primaries; the dashed line represents secondaries. }
\end{figure}

\begin{figure}
\epsscale{1.0}
\plotone{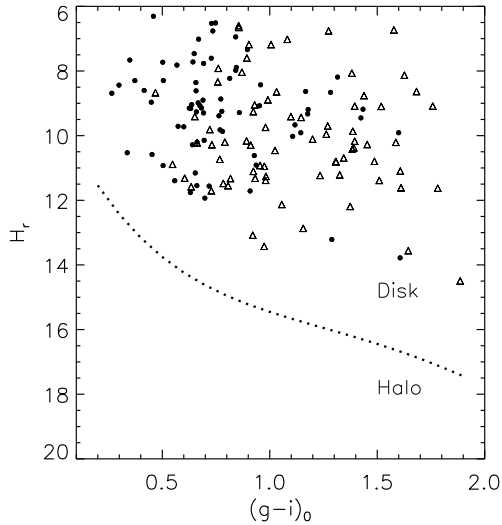}
\caption{RPM of these fragile pairs. Filled circles and open triangles represent the (brighter) primaries and (fainter) secondaries of the pairs, respectively. The dotted line indicates the division between the halo and disk. The definition of this division line is illustrated in Sec. 3.  }
\end{figure}

\begin{figure}
\epsscale{1.0}
\plotone{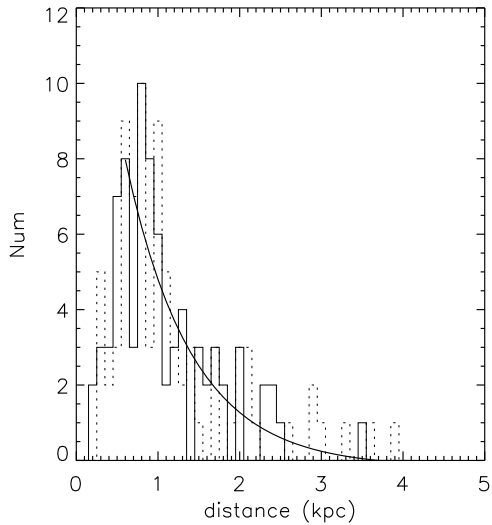}
\caption{Distance distribution of fragile pairs. The solid line represents primaries; the dashed line represents secondaries. Thick solid line is a fit with an exponential law of scale height 0.81 kpc.}
\end{figure}

\begin{figure}
\epsscale{1.0}
\plotone{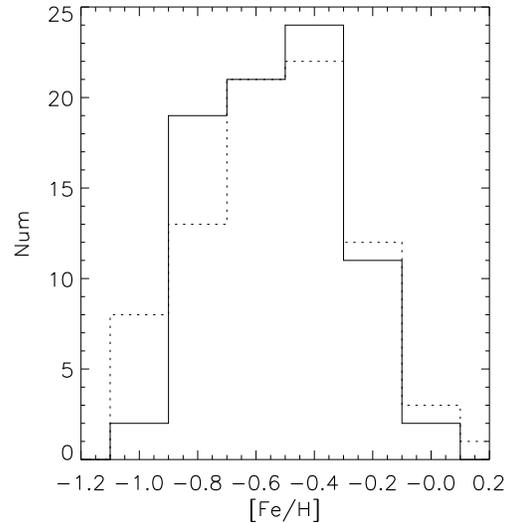}
\caption{[Fe/H] distribution of our pairs. The solid line and dashed line represent primaries and secondaries, respectively.}
\end{figure}

\begin{figure}
\epsscale{1.0}
\plotone{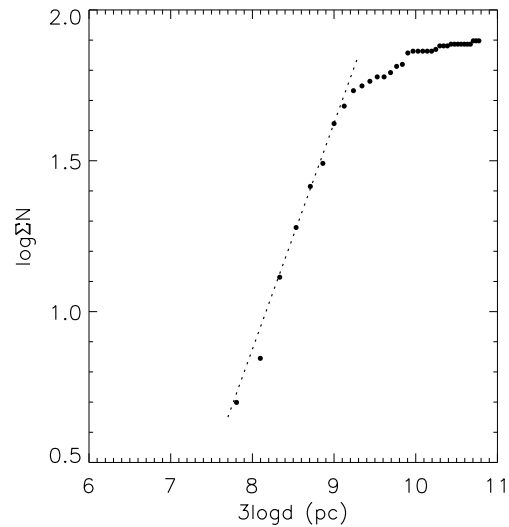}
\caption{Completeness of our {\bf 80 candidate} pairs. {\bf The distance d used for each binary candidate is the average of its two components' distance estimate.} The straight line corresponds to $\Sigma$N(d) $\sim$ d$^{3}$, i.e., a volume-complete sample. }
\end{figure}

\begin{figure}
\epsscale{1.0}
\plotone{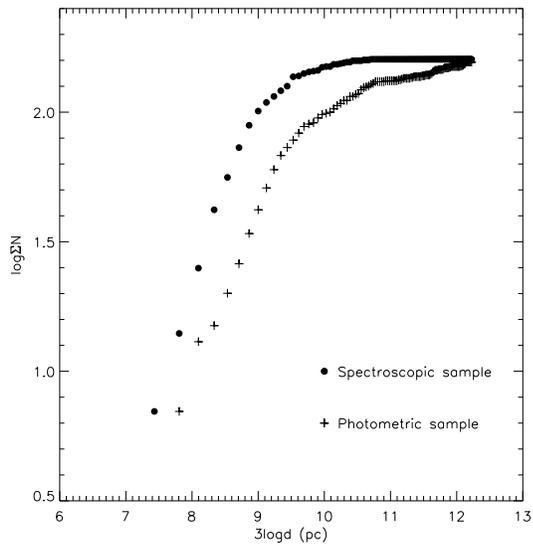}
\caption{Completeness comparison between {\bf 160 stars in our final 80 fragile binary candidates} (filled circles) and 160 stars randomly selected from photometric sample (plus signs). }
\end{figure}

\begin{figure}
\epsscale{1.0}
\plotone{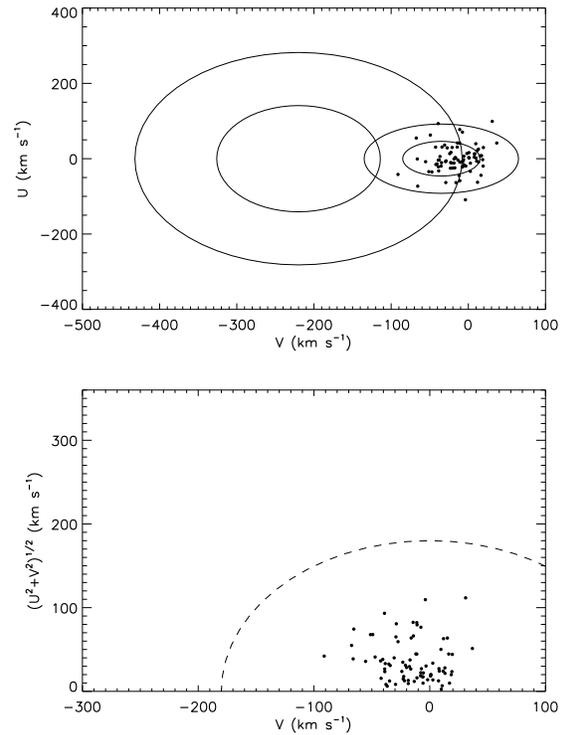}
\caption{ Top: UV-velocity distribution of our pairs. Ellipsoids indicate the 1$\sigma$ (inner) and 2$\sigma$ (outer) contours
for Galactic thick disk and halo populations, respectively. Bottom: Toomre diagram of our pairs. Dashed line
is Vtotal = 180 km s$^{-1}$.}
\end{figure}


\begin{figure}
\epsscale{0.8}
\plotone{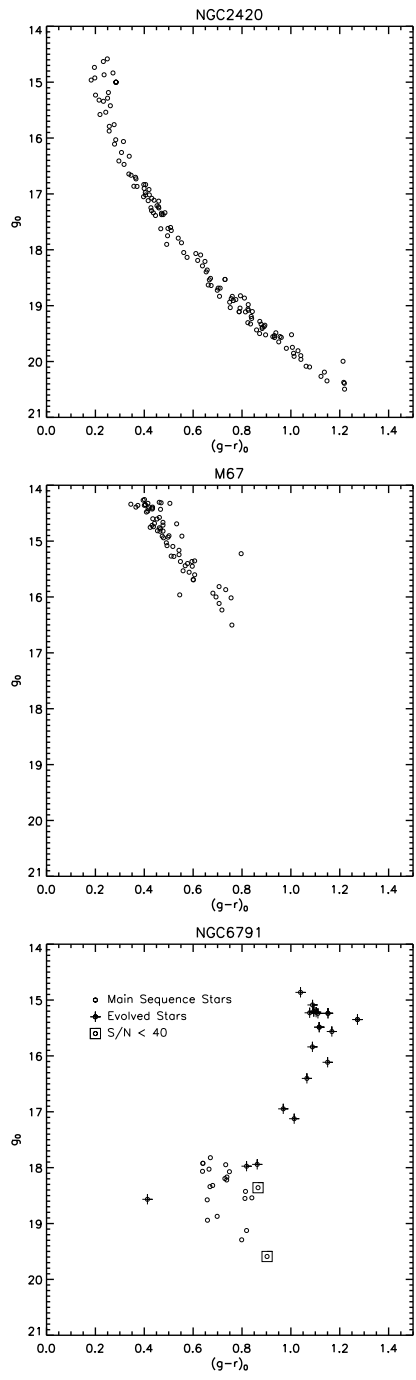}
\caption{The CMD diagrams of NGC2420, M67 and NGC6791.}
\end{figure}

\begin{figure}
\epsscale{1.0}
\plotone{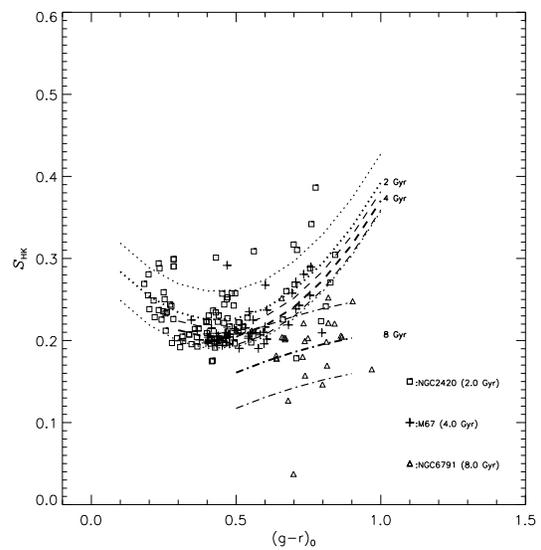}
\caption{$S$$\rm_{HK}$ vs. (g-r)$_{0}$ for three open clusters. Plus signs represent member stars of M67; Squares are stars of NGC2420 while triangles indicate the member stars of NGC6791. The
thick dotted lines are the least squares fitting of NGC2420 and $\pm 1\sigma$; the dashed lines are for M67 while the dashed dot lines are for NGC6791.  }
\end{figure}

\begin{figure}
\epsscale{1.0}
\plotone{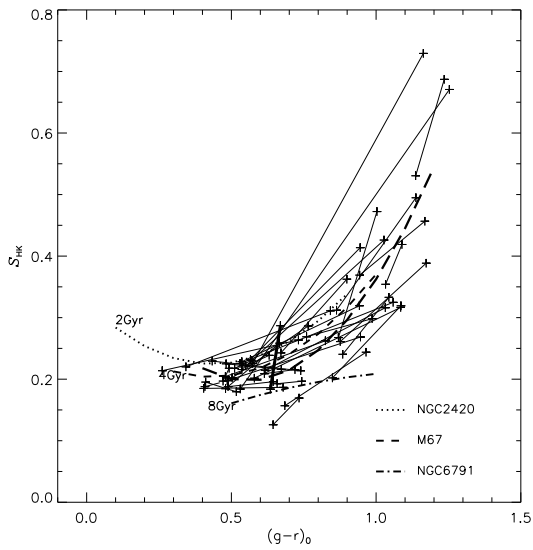}
\caption{$S$$\rm_{HK}$ vs. (g-r)$_{0}$ for fragile binary candidates. Solid lines connect two components of each pair. The dashed line is the distribution of M67 (4.0 Gyr);
 the dotted line are a least fitting of NGC 2420 (2.0 Gyr) while the dash dot line is for NGC6791 (8.0 Gyr). Thick solid line connects the pair which is not a physical pair with high probability. Long dash line is the least squares fitting for all fragile candidates. }
\end{figure}

\clearpage








\clearpage


\begin{deluxetable}{ccccccccccccccccccc}
\tablecolumns{18}
\tabletypesize{\scriptsize}

\setlength{\tabcolsep}{0.01in}
\tablewidth{-1pt}
\tablecaption{properties of fragile binary candidates.}
\tablehead{
\colhead{No}&\colhead{SDSS Obj$_{1}$}    &     \colhead{SDSS Obj$_{2}$}   &
\colhead{pm$\alpha$$_{1}$} & \colhead{pm$\alpha$$_{2}$}   & \colhead{pm$\delta$$_{1}$}    & \colhead{pm$\delta$$_{2}$} &
\colhead{RV$_{1}$}    & \colhead{RV$_{2}$}   & \colhead{[Fe/H]$_{1}$}    & \colhead{[Fe/H]$_{2}$}& \colhead{r$_{1}$}    & \colhead{r$_{2}$}&
 \colhead{(g-i)$_{1}$}   & \colhead{(g-i)$_{2}$}& \colhead{(SP)$_{1}$}& \colhead{(SP)$_{2}$}& \colhead{$\Delta\theta$($\arcmin\arcmin$)}    & \colhead{a(pc)}}
\startdata
1&J141918.7+002550  &J141920.5+002535  & -10.31  &  -8.53  &  -2.85  &  -7.84  & -28.29  & -31.70  &  -0.54  &  -0.78  &  16.79  &  18.11  &   0.70  &   0.97  & F9  & K1  &  31.03  &  0.349\\
2&J004839.6+151923  &J004841.9+151818  &   3.30  &   5.78  &  -3.56  &  -4.91  & -78.84  & -96.16  &  -0.55  &  -0.32  &  14.34  &  17.23  &   0.69  &   1.78  & F9  & K7  &  73.59  &  0.269\\
3&J014701.4+150020  &J014701.8+150014  &  -4.38  &  -7.16  &  -1.68  &  -4.20  &  -4.42  &  -6.53  &  -0.68  &  -0.76  &  16.37  &  18.27  &   0.60  &   1.16  & G2  & K3  &   8.41  &  0.088\\
4&J035711.1$-$070548  &J035715.4$-$070604  &   3.61  &   7.32  &   4.21  &   7.20  &   3.69  & -15.87  &  -0.74  &  -0.90  &  13.94  &  16.33  &   0.35  &   0.98  & F5  & K3  &  65.93  &  0.362\\
5&J100059.9$-$002450  &J100100.4$-$002410  &  -9.24  &  -6.21  &   1.52  &   4.69  &  65.11  &  48.35  &  -0.32  &  -0.61  &  16.69  &  17.03  &   0.66  &   0.78  & F9  & F9  &  40.90  &  0.438\\
6&J094018.8+521326  &J094010.1+521252  &   1.97  &   1.26  &  -6.84  &  -5.96  &  26.56  &  25.53  &  -0.36  &  -0.11  &  14.78  &  15.49  &   0.64  &   0.65  & F9  & F9  &  87.53  &  0.470\\
7&J234742.9$-$001205  &J234741.3$-$001324  &  11.22  &   6.99  &  -5.10  &  -1.02  &  -1.36  &  16.22  &  -0.82  &  -0.72  &  16.11  &  16.57  &   0.72  &   1.31  & F9  & K5  &  83.22  &  0.405\\
8&J004830.1$-$000933  &J004829.7$-$000917  &  11.49  &  14.65  & -17.11  & -16.71  &  15.48  &  17.59  &  -0.66  &  -0.93  &  16.64  &  17.76  &   1.29  &   1.89  & K5  & K7  &  16.99  &  0.072\\
9&J125817.5+590842  &J125817.3+590907  &  -1.95  &  -3.62  &  -2.87  &  -1.97  &   2.67  &   2.71  &  -0.05  &  -0.24  &  14.76  &  16.63  &   0.65  &   1.27  & F9  & K5  &  25.34  &  0.153\\
10&J032230.9$-$004149  &J032231.8$-$004141  &   2.79  &   0.87  &  -2.50  &  -0.57  &   6.08  &  -2.89  &  -0.22  &  -0.12  &  15.43  &  16.94  &   0.37  &   1.08  & F5  & K3  &  16.42  &  0.152\\
11&J031344.0+005201  &J031342.9+005130  &  -1.02  &  -0.42  &  -3.32  &  -0.33  & -20.69  & -21.19  &  -0.45  &  -0.33  &  17.17  &  18.08  &   0.78  &   1.58  & F9  & K7  &  35.36  &  0.316\\
12&J074759.7+183505  &J074800.5+183502  &   1.43  &  -3.81  &  -0.24  &  -1.51  &  60.75  &  58.73  &  -0.51  &  -0.63  &  18.47  &  18.64  &   0.66  &   0.73  & F9  & F9  &  11.66  &  0.283\\
13&J003551.9+004254  &J003550.1+004254  &  -1.76  &  -2.02  & -10.29  &  -8.93  & -25.75  & -26.66  &  -0.39  &  -0.42  &  15.83  &  16.77  &   0.50  &   0.63  & G2  & F9  &  27.50  &  0.314\\
14&J203521.2+761923  &J203515.4+761820  &  -5.33  &  -1.96  &  -8.59  &  -8.72  & -40.60  & -57.97  &  -0.50  &  -0.34  &  14.69  &  16.35  &   0.57  &   0.92  & F9  & K3  &  65.71  &  0.360\\
15&J033702.2$-$004010  &J033709.8$-$004010  &   5.02  &  -0.41  &  -7.72  & -10.18  &  37.63  &  37.34  &  -0.25  &  -0.44  &  14.08  &  14.91  &   0.69  &   1.26  & F9  & K5  & 113.45  &  0.302\\
16&J221941.2+003400  &J221939.4+003412  &  -0.27  &   3.65  &  -4.69  &  -7.85  & -59.01  & -39.95  &  -0.58  &  -0.80  &  16.46  &  17.44  &   0.77  &   1.06  & K1  & K3  &  30.04  &  0.257\\
17&J221941.4$-$000353  &J221938.5$-$000402  &  10.01  &   9.42  & -10.92  &  -9.11  &  38.54  &  27.15  &  -0.51  &  -0.61  &  17.93  &  17.97  &   1.61  &   1.64  & K5  & K7  &  44.38  &  0.312\\
18&J221716.1$-$000346  &J221717.6$-$000315  &   2.46  &   3.99  &  -2.74  &  -2.85  & -19.78  &  -9.28  &  -0.23  &   0.02  &  15.40  &  15.81  &   0.81  &   0.92  & K1  & K1  &  37.82  &  0.217\\
19&J004731.1$-$004620  &J004733.3$-$004607  &   0.95  &   1.62  &  -8.32  &  -5.23  &  -0.58  &   6.18  &  -0.75  &  -0.94  &  16.54  &  17.70  &   0.65  &   1.51  & F9  & K5  &  34.91  &  0.219\\
20&J031142.0$-$005018  &J031142.3$-$005026  &   6.03  &   5.98  &   3.10  &  -0.41  & -22.94  & -29.61  &  -0.69  &  -0.56  &  16.43  &  17.43  &   0.45  &   0.60  & F2  & F9  &   8.82  &  0.133\\
21&J025119.7$-$001345  &J025121.7$-$001317  &   4.19  &   3.27  &  -0.23  &   4.07  &  -8.84  & -11.39  &  -0.19  &  -0.28  &  15.18  &  16.63  &   0.50  &   0.66  & F9  & F9  &  41.39  &  0.383\\
22&J005338.9+000230  &J005340.3+000054  &   5.93  &   2.04  &  -4.91  &  -2.30  &  25.64  &  21.52  &  -0.36  &  -0.63  &  15.48  &  16.65  &   1.60  &   1.76  & K7  & K7  &  98.08  &  0.255\\
23&J030240.2+001000  &J030239.3+000957  &   3.33  &   3.07  &   4.60  &  -0.25  &  40.12  &  24.47  &  -0.62  &  -0.74  &  16.76  &  18.29  &   0.34  &   0.77  & F5  & K1  &  14.77  &  0.275\\
24&J012930.3+402816  &J012922.6+402831  &   4.25  &   6.84  &  -7.36  &  -2.63  &  -2.43  &   0.30  &  -0.10  &  -0.36  &  14.64  &  15.42  &   0.86  &   0.98  & K1  & K1  &  88.79  &  0.362\\
25&J180746.3+243637  &J180748.6+243525  &  -4.54  &  -6.67  &  -4.25  &  -7.83  &  -4.83  &  -9.33  &  -0.63  &  -0.45  &  15.19  &  16.27  &   0.63  &   0.82  & F9  & K1  &  78.24  &  0.457\\
26&J224438.4+230709  &J224433.9+230653  &  -8.07  &  -6.18  &  -7.80  &  -8.01  & -32.38  & -33.39  &  -0.24  &   0.06  &  15.03  &  15.14  &   0.64  &   0.89  & F9  & K1  &  63.77  &  0.341\\
27&J020358.5$-$003207  &J020401.8$-$003233  &   2.35  &   2.00  &  -1.07  &  -2.86  &  18.38  &  26.77  &  -0.77  &  -0.90  &  15.76  &  16.18  &   0.57  &   0.99  & F2  & K1  &  54.84  &  0.279\\
28&J173137.4+333408  &J173136.1+333404  &  -0.31  &  -1.83  & -10.84  & -10.23  &  40.74  &  48.22  &  -0.65  &  -0.83  &  16.22  &  18.00  &   0.56  &   0.92  & F9  & K1  &  16.80  &  0.181\\
29&J012439.9+402031  &J012441.9+402013  &  -4.08  &  -3.12  &  -0.52  &  -4.50  & -25.90  & -27.81  &  -0.80  &  -0.78  &  15.37  &  17.20  &   0.30  &   0.55  & F5  & F9  &  29.22  &  0.425\\
30&J024416.0+004725  &J024417.9+004719  &   0.53  &  -1.73  &   3.20  &   5.58  &  13.55  &  11.90  &  -0.24  &  -0.40  &  15.17  &  17.78  &   0.64  &   1.61  & F9  & K7  &  28.96  &  0.194\\
31&J024604.3+011348  &J024601.7+011348  &  -0.37  &  -0.34  &   1.92  &   0.79  &  57.03  &  69.64  &  -0.78  &  -0.59  &  14.85  &  17.08  &   0.46  &   1.27  & F5  & K5  &  38.69  &  0.260\\
32&J040921.1$-$052701  &J040927.0$-$052655  &   6.83  &   4.64  &  -3.04  &   1.29  &  58.14  &  53.12  &  -0.62  &  -0.46  &  15.02  &  16.80  &   0.76  &   1.59  & F9  & K7  &  89.43  &  0.390\\
33&J124144.1$-$014829  &J124148.1$-$014952  &  -5.77  &  -7.30  &   6.39  &   8.68  &   4.79  &   6.88  &  -0.44  &  -0.40  &  14.31  &  15.67  &   0.67  &   0.97  & F9  & K3  & 102.22  &  0.404\\
34&J123925.0$-$023812  &J123925.2$-$023739  &  -6.24  &  -2.19  &  -6.80  &  -5.44  &  45.87  &  50.03  &  -0.49  &  -0.58  &  14.47  &  16.02  &   0.69  &   1.39  & F9  & K5  &  32.59  &  0.122\\
35&J031737.1$-$072658  &J031738.0$-$072533  &   0.62  &   1.77  &   3.70  &   8.30  &  33.16  &  31.09  &  -0.79  &  -0.91  &  16.21  &  16.28  &   0.95  &   0.95  & K1  & K1  &  86.36  &  0.476\\
36&J003044.2+142906  &J003040.7+142921  &   3.30  &   3.53  &   3.30  &   0.09  &   1.95  & -11.30  &  -0.06  &  -0.02  &  15.29  &  16.35  &   1.17  &   1.39  & K5  & K5  &  53.14  &  0.218\\
37&J095600.5+002626  &J095601.4+002551  &  -5.14  &  -0.25  &  -1.67  &  -0.18  &  -7.95  & -14.07  &  -0.40  &  -0.52  &  15.41  &  15.90  &   0.67  &   1.05  & F9  & K3  &  37.29  &  0.183\\
38&J170531.0+364758  &J170535.9+364818  &  -3.98  &  -1.54  &   4.43  &   3.15  & -33.18  & -27.41  &  -0.20  &  -0.42  &  15.27  &  16.69  &   0.63  &   1.10  & F9  & K3  &  63.05  &  0.443\\
39&J082116.6+374008  &J082122.1+374055  &  -4.38  &  -0.24  &  -3.01  &  -2.29  & -24.07  & -30.74  &  -0.29  &  -0.25  &  16.04  &  16.95  &   1.12  &   1.44  & K3  & K5  &  80.74  &  0.438\\
40&J082923.4+394705  &J082929.6+394635  &   1.38  &  -3.67  &  -2.73  &  -2.07  &  11.81  &  -0.05  &  -0.62  &  -0.37  &  15.55  &  17.05  &   0.84  &   1.39  & F9  & K5  &  77.95  &  0.390\\
41&J082555.4+384633  &J082602.0+384723  &  -3.08  &  -5.86  &   0.91  &   0.80  &   6.55  &  23.15  &  -0.74  &  -0.60  &  15.90  &  16.94  &   0.96  &   1.31  & F9  & K5  &  92.40  &  0.453\\
42&J081940.2+320117  &J081939.3+320049  &   1.82  &   3.16  &  -1.18  &   3.64  &  69.53  &  50.02  &  -0.11  &  -0.37  &  16.98  &  17.28  &   1.28  &   1.34  & K5  & K5  &  31.34  &  0.238\\
43&J092513.8+442356  &J092519.3+442251  &  -1.57  &   3.99  &   3.64  &   1.73  & -20.43  &  -9.62  &  -0.36  &  -0.40  &  16.20  &  17.09  &   1.18  &   1.45  & K3  & K5  &  87.69  &  0.471\\
44&J075253.3+282215  &J075250.4+282241  &  -2.50  &  -5.43  &  -5.62  &  -1.16  &  40.64  &  35.38  &  -0.37  &  -0.30  &  16.67  &  17.54  &   0.93  &   0.98  & K1  & K3  &  46.83  &  0.396\\
45&J141146.0+455747  &J141152.8+455937  &  -0.90  &  -0.36  &  -6.06  &  -3.66  &  -0.40  &  -5.93  &  -0.13  &  -0.34  &  15.25  &  16.27  &   1.43  &   1.52  & K5  & K7  & 130.22  &  0.400\\
46&J131044.4+502744  &J131039.9+502837  &  -0.41  &  -4.02  &  -4.34  &  -3.75  & -41.40  & -37.63  &  -0.96  &  -0.66  &  15.67  &  17.09  &   0.77  &   1.49  & F9  & K5  &  68.52  &  0.350\\
47&J102043.7+094304  &J102043.8+094151  &  -0.74  &  -2.08  &  -4.94  &  -1.62  &  16.51  &   9.43  &  -0.62  &  -0.43  &  16.53  &  17.33  &   1.11  &   1.15  & K3  & K3  &  73.13  &  0.435\\
48&J103034.7+091759  &J103034.3+091805  &  -3.40  &  -2.80  &   3.97  &   6.21  &  33.99  &  28.85  &  -0.42  &  -0.36  &  16.55  &  17.38  &   0.69  &   0.81  & F9  & F9  &   9.38  &  0.100\\
49&J100705.5+121349  &J100701.7+121312  &   4.27  &  -1.08  &   6.89  &   5.68  &  45.59  &  43.93  &  -0.79  &  -0.89  &  16.37  &  17.40  &   0.94  &   1.33  & K3  & K5  &  67.13  &  0.410\\
50&J074435.9+175738  &J074434.8+175727  &  -5.82  &  -2.79  &  -0.72  &   0.98  &  70.44  &  86.97  &  -0.59  &  -0.58  &  14.85  &  17.93  &   0.27  &   0.73  & F5  & F9  &  19.62  &  0.277\\
51&J130109.3+493750  &J130115.9+493843  &   0.19  &  -3.13  &   3.09  &  -0.35  &  -4.89  & -14.18  &  -0.41  &  -0.15  &  15.74  &  16.15  &   1.32  &   1.68  & K5  & K7  &  82.99  &  0.297\\
52&J143126.9+081320  &J143122.6+081311  & -12.07  &  -9.05  &   0.79  &  -0.10  & -20.12  &  -2.85  &  -0.71  &  -0.81  &  16.35  &  16.54  &   0.63  &   0.93  & F9  & K3  &  66.15  &  0.436\\
53&J145450.7+105620  &J145446.7+105714  &  -4.50  &  -1.44  &  -3.14  &  -7.07  &   3.15  &  -0.42  &  -0.42  &  -0.14  &  15.63  &  16.12  &   1.18  &   1.38  & K3  & K5  &  80.14  &  0.323\\
54&J145257.7+325152  &J145300.8+325112  &   6.16  &   3.38  &  -6.08  &  -4.27  & -47.01  & -45.70  &  -0.51  &  -0.68  &  17.03  &  17.55  &   0.91  &   1.23  & K1  & K5  &  56.22  &  0.467\\
55&J221556.6+682321  &J221607.4+682039  &   4.28  &   6.57  &  -2.71  &   1.27  & -47.57  & -48.26  &  -0.28  &  -0.40  &  13.82  &  14.52  &   0.90  &   1.03  & K3  & K5  & 172.83  &  0.423\\
56&J221751.8+690949  &J221726.7+690953  &  -0.86  &   0.23  &  -2.90  &  -5.10  & -50.46  & -64.71  &  -0.35  &  -0.22  &  14.12  &  14.80  &   0.73  &   0.76  & K1  & K1  & 133.97  &  0.457\\
57&J221428.5+682529  &J221415.1+682619  &  -0.51  &   2.11  &   1.93  &   1.04  &-102.11  &-114.81  &  -0.41  &  -0.31  &  14.29  &  14.80  &   0.74  &   0.86  & K3  & K3  &  89.31  &  0.313\\
58&J213425.9+730017  &J213424.9+725850  &  -0.60  &  -0.14  &  -3.50  &   0.29  & -18.34  &  -2.91  &  -0.39  &  -0.34  &  14.01  &  14.40  &   0.74  &   0.84  & K3  & G5  &  87.19  &  0.275\\
59&J213410.2+734401  &J213423.2+734547  &   3.66  &   1.13  &   2.38  &   6.20  & -40.01  & -38.08  &  -0.23  &  -0.43  &  14.69  &  15.05  &   0.84  &   0.93  & K3  & K3  & 119.35  &  0.465\\
60&J213206.3+750645  &J213148.2+750553  &  -1.35  &  -0.21  &   0.41  &  -3.95  & -12.15  &   1.13  &  -0.07  &  -0.19  &  13.95  &  14.61  &   0.82  &   0.89  & K1  & K3  &  87.13  &  0.274\\
61&J010418.2+002633  &J010422.3+002755  &  -0.05  &  -1.26  &  -0.85  &  -1.52  &  -1.06  &  -5.70  &  -0.39  &  -0.39  &  15.59  &  16.59  &   1.01  &   1.38  & K3  & K5  & 102.77  &  0.478\\
62&J093741.6+291905  &J093741.9+291737  &  -8.11  &  -6.30  &   2.62  &   0.12  &   6.93  &   8.56  &  -0.55  &  -0.60  &  14.31  &  16.11  &   0.45  &   1.20  & G0  & K3  &  88.21  &  0.402\\
63&J161348.9+505831  &J161352.5+505729  &  -4.76  &  -1.62  &   4.70  &   5.64  & -32.02  & -12.83  &  -0.23  &  -0.15  &  15.78  &  17.25  &   1.14  &   1.61  & K3  & K7  &  70.69  &  0.342\\
64&J073427.2+652057  &J073427.8+652037  &   1.06  &   0.40  &   0.85  &  -4.33  &  10.44  &  12.68  &  -0.44  &  -0.56  &  16.35  &  17.01  &   0.67  &   0.79  & F9  & F9  &  20.73  &  0.209\\
65&J080736.9+664653  &J080725.3+664722  &   3.80  &   6.01  &  -3.40  &   0.67  & -18.24  & -14.37  &  -0.55  &  -0.60  &  15.71  &  16.38  &   0.78  &   0.91  & F9  & K1  &  74.64  &  0.452\\
66&J191817.1+365335  &J191812.9+365322  &   4.95  &   3.53  &  -0.28  &  -4.30  &   4.42  & -14.68  &  -0.22  &  -0.45  &  15.67  &  18.47  &   0.68  &   1.37  & F9  & K5  &  52.02  &  0.414\\
67&J201548.9$-$130112  &J201552.6$-$130114  &  -3.27  &  -1.50  &  -0.57  &  -0.84  & -16.13  & -17.80  &  -0.17  &  -0.20  &  16.85  &  16.95  &   1.42  &   1.63  & K5  & K7  &  53.68  &  0.288\\
68&J023335.5+264149  &J023337.4+264204  &  -1.66  &   0.92  &  -1.09  &   0.20  & -15.36  & -12.47  &  -0.63  &  -0.53  &  16.24  &  17.32  &   0.50  &   0.90  & G2  & K1  &  29.05  &  0.317\\
69&J044715.1+214438  &J044715.8+214254  &   2.25  &   2.86  &  -2.58  &  -3.53  &  24.27  &  23.79  &  -0.34  &  -0.45  &  13.84  &  14.64  &   0.75  &   0.76  & K1  & K1  & 104.24  &  0.305\\
70&J204024.3+561458  &J204003.0+561354  &  -1.69  &  -2.85  &  -1.27  &   1.26  & -47.91  & -40.59  &  -0.03  &  -0.08  &  13.01  &  13.40  &   0.87  &   1.05  & K3  & K5  & 188.62  &  0.360\\
71&J204212.9+560534  &J204228.2+560539  &   1.15  &  -2.16  &   2.73  &   0.98  & -34.86  & -51.14  &  -0.15  &  -0.00  &  12.89  &  13.51  &   0.84  &   0.94  & K3  & K5  & 127.29  &  0.232\\
72&J201647.3+595717  &J201650.4+595717  &  -6.50  &  -5.67  &   2.10  &  -3.82  & -35.71  & -35.68  &   0.05  &  -0.09  &  14.44  &  16.20  &   0.66  &   1.39  & F9  & K5  &  23.16  &  0.117\\
73&J205345.0+573901  &J205334.8+574102  &  -1.00  &   4.90  &   1.36  &   2.42  & -23.77  & -42.80  &  -0.35  &  -0.06  &  13.10  &  13.50  &   0.85  &   1.01  & K3  & K5  & 145.45  &  0.262\\
74&J192106.3+374460  &J192058.8+374313  &   0.46  &   1.84  &   0.02  &  -1.90  & -45.81  & -45.27  &   0.46  &   0.27  &  13.59  &  13.70  &   1.19  &   1.23  & K5  & K5  & 139.18  &  0.308\\
75&J203959.7+564719  &J203941.7+564526  &  -1.34  &  -1.78  &   1.24  &   6.92  & -26.99  & -10.22  &  -0.42  &  -0.40  &  13.19  &  13.77  &   0.75  &   0.87  & K5  & K5  & 186.16  &  0.386\\
76&J052006.9+172819  &J052013.5+172714  &   3.04  &   1.90  &   1.39  &   1.09  &  25.22  &  19.46  &  -0.22  &  -0.42  &  14.32  &  14.89  &   0.84  &   0.86  & K1  & K1  & 114.11  &  0.388\\
77&J042405.9+070725  &J042402.0+070717  &   8.97  &   5.54  &  -4.44  &  -6.15  &   1.23  &   0.77  &  -0.35  &  -0.49  &  15.20  &  15.22  &   0.66  &   0.72  & K1  & F9  &  59.47  &  0.321\\
78&J051613.4+165303  &J051617.6+165145  &  -1.06  &  -0.83  &  -3.75  &   0.36  &  56.69  &  49.23  &   0.06  &  -0.03  &  14.65  &  14.67  &   0.73  &   0.91  & K1  & K1  &  98.20  &  0.397\\
79&J063503.6+275149  &J063503.9+275139  &  -0.16  &  -0.20  &   2.35  &   1.83  &  50.83  &  35.96  &  -0.41  &  -0.68  &  16.74  &  17.35  &   0.42  &   0.47  & G2  & G0  &  11.50  &  0.251\\
80&J072816.6+145419  &J072815.2+145414  &  -0.56  &   1.11  &   1.58  &   1.82  &  82.15  &  84.99  &  -0.42  &  -0.34  &  17.24  &  18.81  &   0.66  &   1.02  & F9  & K1  &  20.91  &  0.328\\

\enddata
\end{deluxetable}


\clearpage
\begin{deluxetable}{cccccccccccccccc}
\tablecolumns{18}
\tabletypesize{\scriptsize}

\setlength{\tabcolsep}{0.05in}
\tablewidth{-1pt}
\tablecaption{properties of fragile binary candidates.}
\tablehead{
\colhead{No}&\colhead{U$_{1}$}  &     \colhead{V$_{1}$}   &
\colhead{W$_{1}$} & \colhead{U$_{2}$}  &     \colhead{V$_{2}$}   &
\colhead{W$_{2}$} &
 \colhead{dist$_{1}$}    & \colhead{dist$_{2}$}&
 \colhead{$\Delta$U}   & \colhead{$\Delta$V}& \colhead{$\Delta$W}& \colhead{$\Delta$dist}& \colhead{age consistency}& \colhead{confidence}\\
&&(km s$^{-1}$)&&&(km s$^{-1}$)&&(pc)&
(pc)& &(km s$^{-1}$)&&(pc)&\\}
\startdata
 1&    54$\pm$42&   -67$\pm$48&     2$\pm$27&    27$\pm$58&  -106$\pm$67&   -26$\pm$40&   1841&   2147&    27&    39&    28&    306&  NULL  &E   \\
 2&   -34$\pm$12&   -51$\pm$11&    58$\pm$ 5&   -33$\pm$18&   -72$\pm$17&    65$\pm$ 8&    598&    789&     1&    21&     7&    191&    Y & C   \\
 3&   -44$\pm$28&    16$\pm$32&    -6$\pm$19&   -77$\pm$42&    18$\pm$47&   -28$\pm$29&   1711&   1969&    33&     2&    22&    258&  NULL   &E  \\
 4&    11$\pm$17&     7$\pm$23&    24$\pm$16&     8$\pm$17&    10$\pm$23&    49$\pm$16&   1056&    899&     3&     3&    25&    157&   Y& C   \\
 5&    93$\pm$45&   -39$\pm$28&     1$\pm$38&    67$\pm$38&    -4$\pm$26&    25$\pm$32&   2065&   1753&    26&    35&    24&    312&   NULL&  E \\
 6&     3$\pm$10&   -17$\pm$15&    37$\pm$11&     3$\pm$15&   -26$\pm$23&    38$\pm$16&    878&   1333&     0&     9&     1&    455&     NULL &B  \\
 7&    30$\pm$28&   -42$\pm$23&   -19$\pm$12&    11$\pm$16&     0$\pm$13&   -14$\pm$ 6&   1165&    796&    19&    42&     5&    369&    Y&  E  \\
 8&     0$\pm$25&   -66$\pm$25&   -38$\pm$ 8&     8$\pm$23&   -54$\pm$23&   -33$\pm$ 6&    860&    688&     8&    12&     5&    172&    NULL & C  \\
 9&    -8$\pm$19&    -8$\pm$18&    16$\pm$ 9&     0$\pm$20&   -10$\pm$20&    15$\pm$ 9&    988&   1042&     8&     2&     1&     54&    Y& A   \\
10&    -3$\pm$33&   -37$\pm$46&     7$\pm$31&   -10$\pm$21&    -1$\pm$28&    10$\pm$18&   2428&   1511&     7&    36&     3&    917&     NULL &  E \\
11&   -44$\pm$30&   -12$\pm$40&     2$\pm$26&   -27$\pm$24&     5$\pm$31&    19$\pm$20&   2010&   1462&    17&    17&    17&    548&   NULL  & D  \\
12&    28$\pm$51&   -27$\pm$84&    51$\pm$92&    66$\pm$51&   -24$\pm$83&   -44$\pm$92&   4319&   3973&    38&     3&    95&    346&    NULL  & E \\
13&   -72$\pm$37&   -65$\pm$39&   -14$\pm$16&   -76$\pm$43&   -63$\pm$44&   -10$\pm$18&   1903&   2154&     4&     2&     4&    251&      Y& E \\
14&   -63$\pm$20&   -16$\pm$10&    -6$\pm$19&   -72$\pm$28&   -23$\pm$14&   -28$\pm$26&    897&   1214&     9&     7&    22&    317&    NULL  & D \\
15&    13$\pm$10&   -24$\pm$14&   -19$\pm$ 9&     6$\pm$ 7&   -12$\pm$ 9&   -27$\pm$ 5&    611&    435&     7&    12&     8&    176&      Y&  A \\
16&    -7$\pm$25&   -55$\pm$20&    34$\pm$18&    -3$\pm$32&   -65$\pm$26&    -3$\pm$25&   1398&   1448&     4&    10&    37&     50&    NULL  & D \\
17&    -5$\pm$34&   -28$\pm$26&   -80$\pm$25&     0$\pm$32&   -24$\pm$25&   -62$\pm$21&   1224&   1151&     5&     4&    18&     73&    NULL  & D \\
18&     0$\pm$17&   -18$\pm$12&     9$\pm$12&     2$\pm$20&   -16$\pm$15&    -5$\pm$16&    940&   1124&     2&     2&    14&    184&    Y& A   \\
19&   -35$\pm$31&   -47$\pm$33&   -20$\pm$11&   -14$\pm$21&   -15$\pm$22&    -9$\pm$ 5&   1627&   1026&    21&    32&    11&    601&    NULL  & E \\
20&    30$\pm$39&   -14$\pm$51&    76$\pm$33&    11$\pm$45&   -53$\pm$58&    68$\pm$38&   2475&   2904&    19&    39&     8&    429&     NULL  & E \\
21&     1$\pm$24&   -16$\pm$29&    26$\pm$18&    21$\pm$35&    13$\pm$44&    47$\pm$27&   1515&   2044&    20&    29&    21&    529&   NULL  &  E \\
22&     1$\pm$ 9&     0$\pm$ 9&   -20$\pm$ 1&    -2$\pm$10&     6$\pm$10&   -14$\pm$ 2&    426&    544&     3&     6&     6&    118&      Y& A  \\
23&    99$\pm$71&    30$\pm$93&    51$\pm$60&    29$\pm$58&   -26$\pm$75&     8$\pm$48&   4224&   3043&    70&    56&    43&   1181&    NULL &  E \\
24&    -3$\pm$ 9&   -13$\pm$10&   -11$\pm$11&     7$\pm$11&   -12$\pm$11&     2$\pm$10&    667&    738&    10&     1&    13&     71&     Y& A  \\
25&   -24$\pm$13&   -17$\pm$14&    16$\pm$17&   -45$\pm$20&   -43$\pm$21&    22$\pm$26&    955&   1264&    21&    26&     6&    309&      Y&  E \\
26&   -63$\pm$20&   -29$\pm$13&    14$\pm$20&   -48$\pm$19&   -31$\pm$11&    12$\pm$16&   1036&    875&    15&     2&     2&    161&      Y&  A \\
27&     6$\pm$19&    -5$\pm$20&    -7$\pm$11&     3$\pm$14&    -3$\pm$15&   -18$\pm$ 7&   1332&    833&     3&     2&    11&    499&      Y& A \\
28&  -109$\pm$32&    -3$\pm$27&    10$\pm$26&  -122$\pm$42&   -11$\pm$37&    25$\pm$37&   1766&   2120&    13&     8&    15&    354&    NULL  & E \\
29&   -62$\pm$31&    11$\pm$31&     5$\pm$29&   -65$\pm$38&   -12$\pm$40&   -40$\pm$42&   2381&   2693&     3&    23&    45&    312&    NULL  & E \\
30&     8$\pm$15&    17$\pm$19&     5$\pm$11&     6$\pm$21&    36$\pm$26&     8$\pm$14&   1099&   1194&     2&    19&     3&     95&     NULL   & D\\
31&    29$\pm$15&    19$\pm$19&   -32$\pm$10&    34$\pm$15&    15$\pm$19&   -45$\pm$11&   1135&   1098&     5&     4&    13&     37&    NULL  & B \\
32&    36$\pm$11&   -30$\pm$16&   -17$\pm$13&    38$\pm$12&   -14$\pm$17&   -12$\pm$13&    713&    758&     2&    16&     5&     45&      Y& A  \\
33&    13$\pm$15&     8$\pm$14&    20$\pm$ 6&    28$\pm$20&    12$\pm$19&    28$\pm$ 8&    647&    819&    15&     4&     8&    172&     NULL   &B \\
34&   -13$\pm$15&   -40$\pm$15&    35$\pm$ 7&   -23$\pm$13&   -31$\pm$12&    42$\pm$ 6&    653&    612&    10&     9&     7&     41&       Y& A \\
35&    21$\pm$16&    11$\pm$19&   -11$\pm$10&    33$\pm$18&    22$\pm$21&     0$\pm$11&    912&    902&    12&    11&    11&     10&    NULL  & B \\
36&     4$\pm$13&     7$\pm$11&    11$\pm$ 7&     0$\pm$17&    -9$\pm$15&    14$\pm$ 8&    671&    913&     4&    16&     3&    242&       Y& A \\
37&     3$\pm$20&     0$\pm$13&   -17$\pm$17&   -15$\pm$13&    13$\pm$ 9&    -2$\pm$11&   1076&    804&    18&    13&    15&    272&     Y& C  \\
38&    29$\pm$20&   -20$\pm$19&     9$\pm$18&    17$\pm$19&   -11$\pm$18&     0$\pm$16&   1214&   1151&    12&     9&     9&     63&      Y&  A \\
39&   -19$\pm$ 8&    -3$\pm$14&   -23$\pm$13&   -34$\pm$ 9&    -3$\pm$17&   -12$\pm$14&    888&   1036&    15&     0&    11&    148&     Y& A  \\
40&    -2$\pm$ 7&    -6$\pm$12&    17$\pm$10&     1$\pm$11&    -2$\pm$17&    -9$\pm$15&    820&   1067&     3&     4&    26&    247&     Y&  E \\
41&     1$\pm$ 7&     9$\pm$12&     1$\pm$11&    24$\pm$11&    11$\pm$16&    -1$\pm$16&    802&    993&    23&     2&     2&    191&      Y&  C \\
42&    41$\pm$12&   -14$\pm$21&    51$\pm$20&    22$\pm$13&    15$\pm$23&    53$\pm$21&   1290&   1240&    19&    29&     2&     50&      NULL  & E\\
43&   -19$\pm$ 9&    19$\pm$13&   -12$\pm$ 9&   -30$\pm$12&    13$\pm$15&    13$\pm$13&    879&   1014&    11&     6&    25&    135&     Y& C  \\
44&    30$\pm$12&   -34$\pm$27&     0$\pm$24&    44$\pm$20&     0$\pm$37&   -25$\pm$40&   1382&   2008&    14&    34&    25&    626&    NULL  & E \\
45&   -18$\pm$10&    -5$\pm$ 9&    11$\pm$ 4&   -17$\pm$12&    -5$\pm$11&     5$\pm$ 6&    502&    671&     1&     0&     6&    169&      Y& A  \\
46&   -25$\pm$15&   -23$\pm$15&   -23$\pm$ 6&   -11$\pm$19&   -29$\pm$19&   -19$\pm$ 7&    835&    883&    14&     6&     4&     48&      Y& A  \\
47&   -10$\pm$16&   -22$\pm$15&    10$\pm$12&     0$\pm$26&   -11$\pm$22&     2$\pm$19&    974&   1476&    10&    11&     8&    502&      NULL  & D\\
48&    39$\pm$32&     9$\pm$28&    30$\pm$22&    53$\pm$44&    39$\pm$40&    36$\pm$30&   1749&   2233&    14&    30&     6&    484&       Y& E \\
49&     7$\pm$18&    15$\pm$16&    63$\pm$14&    26$\pm$20&     8$\pm$19&    45$\pm$16&    999&   1081&    19&     7&    18&     82&    NULL  & D \\
50&    77$\pm$24&   -11$\pm$37&   -27$\pm$45&    85$\pm$33&    -4$\pm$54&     7$\pm$61&   2313&   2903&     8&     7&    34&    590&    NULL  & E \\
51&    -6$\pm$11&    10$\pm$11&     0$\pm$ 4&    -5$\pm$11&    -5$\pm$11&    -5$\pm$ 3&    614&    584&     1&    15&     5&     30&       Y& A \\
52&    62$\pm$33&   -49$\pm$37&    25$\pm$20&    21$\pm$21&   -25$\pm$24&    21$\pm$13&   1579&   1078&    41&    24&     4&    501&      Y&  E \\
53&    -9$\pm$11&   -11$\pm$13&    13$\pm$ 7&   -22$\pm$14&   -17$\pm$17&     1$\pm$ 9&    659&    784&    13&     6&    12&    125&      Y& A \\
54&   -58$\pm$37&   -10$\pm$36&   -53$\pm$14&   -31$\pm$31&   -14$\pm$31&   -42$\pm$12&   1564&   1359&    27&     4&    11&    205&     Y&  E \\
55&   -21$\pm$ 8&   -38$\pm$ 4&    -9$\pm$ 8&   -12$\pm$ 9&   -43$\pm$ 4&    -6$\pm$ 9&    400&    454&     9&     5&     3&     54&      NULL  &B \\
56&   -32$\pm$10&   -38$\pm$ 5&    -6$\pm$10&   -39$\pm$13&   -48$\pm$ 6&   -20$\pm$15&    558&    774&     7&    10&    14&    216&    NULL  & B \\
57&   -41$\pm$10&   -91$\pm$ 5&    -5$\pm$10&   -40$\pm$12&  -103$\pm$ 6&   -13$\pm$12&    574&    650&     1&    12&     8&     76&     NULL   & B\\
58&   -22$\pm$ 9&    -7$\pm$ 4&    -2$\pm$10&   -10$\pm$ 9&     2$\pm$ 4&     7$\pm$ 9&    516&    542&    12&     9&     9&     26&     NULL   & B\\
59&   -10$\pm$13&   -34$\pm$ 5&    -6$\pm$13&    -9$\pm$13&   -37$\pm$ 6&     7$\pm$13&    652&    637&     1&     3&    13&     15&    NULL  & B \\
60&   -16$\pm$ 9&    -5$\pm$ 4&     6$\pm$ 9&   -16$\pm$11&    11$\pm$ 5&     0$\pm$11&    515&    605&     0&    16&     6&     90&      N&  E \\
61&   -11$\pm$12&     2$\pm$12&     6$\pm$ 4&   -18$\pm$15&     1$\pm$15&     9$\pm$ 5&    761&    866&     7&     1&     3&    105&       Y& A \\
62&    25$\pm$15&    13$\pm$14&   -11$\pm$13&    12$\pm$10&     2$\pm$ 9&    -1$\pm$ 9&    998&    746&    13&    11&    10&    252&     Y&  A \\
63&    16$\pm$15&   -22$\pm$12&    -5$\pm$10&    20$\pm$22&     0$\pm$16&     0$\pm$14&    792&   1055&     4&    22&     5&    263&      Y& C  \\
64&    -8$\pm$20&    13$\pm$24&    19$\pm$19&    16$\pm$23&   -21$\pm$29&    14$\pm$21&   1653&   1759&    24&    34&     5&    106&      Y& E \\
65&   -24$\pm$14&   -20$\pm$15&    13$\pm$13&   -38$\pm$16&    -3$\pm$16&    25$\pm$17&    991&   1117&    14&    17&    12&    126&    NULL &  D \\
66&    -2$\pm$20&    18$\pm$11&   -19$\pm$24&   -28$\pm$46&    -7$\pm$22&   -42$\pm$49&   1301&   2014&    26&    25&    23&    713&    NULL &  E \\
67&    -6$\pm$12&    -7$\pm$13&    26$\pm$16&     0$\pm$10&    -7$\pm$12&    18$\pm$13&   1040&    879&     6&     0&     8&    161&     NULL  & B \\
68&   -33$\pm$22&     2$\pm$32&     0$\pm$27&   -13$\pm$20&    -4$\pm$30&    17$\pm$26&   2049&   1783&    20&     6&    17&    266&     Y& C  \\
69&    13$\pm$ 3&    -1$\pm$ 9&     1$\pm$ 8&    13$\pm$ 4&    -7$\pm$12&     1$\pm$11&    478&    645&     0&     6&     0&    167&    NULL  & B \\
70&   -15$\pm$ 5&   -42$\pm$ 2&     0$\pm$ 5&   -13$\pm$ 5&   -35$\pm$ 2&     5$\pm$ 5&    322&    312&     2&     7&     5&     10&     NULL  & B \\
71&    -7$\pm$ 5&   -29$\pm$ 2&     3$\pm$ 5&   -13$\pm$ 6&   -45$\pm$ 2&     3$\pm$ 6&    298&    381&     6&    16&     0&     83&       Y& C \\
72&   -19$\pm$16&   -35$\pm$ 5&    25$\pm$17&   -37$\pm$17&   -30$\pm$ 5&     9$\pm$17&    884&    827&    18&     5&    16&     57&      Y& C \\
73&   -11$\pm$ 5&   -18$\pm$ 2&     6$\pm$ 5&    -5$\pm$ 6&   -37$\pm$ 2&    -2$\pm$ 6&    294&    341&     6&    19&     8&     47&     NULL  & D \\
74&     5$\pm$ 5&   -36$\pm$ 2&    -2$\pm$ 6&     3$\pm$ 5&   -36$\pm$ 2&    -5$\pm$ 6&    399&    362&     2&     0&     3&     37&     NULL  & B \\
75&   -11$\pm$ 5&   -21$\pm$ 3&     5$\pm$ 6&    -2$\pm$ 7&    -6$\pm$ 3&    15$\pm$ 7&    339&    382&     9&    15&    10&     43&    NULL & B  \\
76&    16$\pm$ 3&     0$\pm$ 9&    10$\pm$ 9&    10$\pm$ 3&     2$\pm$11&    10$\pm$10&    556&    648&     6&     2&     0&     92&     Y  & A \\
77&    -3$\pm$11&   -38$\pm$22&    26$\pm$19&   -11$\pm$10&   -29$\pm$18&     9$\pm$15&   1023&    882&     8&     9&    17&    141&    NULL & D  \\
78&    41$\pm$ 4&   -11$\pm$15&   -16$\pm$14&    37$\pm$ 3&     2$\pm$10&    -4$\pm$10&    870&    661&     4&    13&    12&    209&    NULL & B  \\
79&    41$\pm$12&    36$\pm$63&    29$\pm$60&    26$\pm$13&    31$\pm$66&    23$\pm$64&   3574&   3642&    15&     5&     6&     68&     Y  & E \\
80&    70$\pm$23&    -7$\pm$40&    29$\pm$42&    66$\pm$38&    -7$\pm$65&    56$\pm$70&   2566&   3399&     4&     0&    27&    833&    NULL &  E \\

\enddata
\end{deluxetable}

\begin{deluxetable}{ccccc}
\tablecolumns{18}
\tabletypesize{\scriptsize}

\setlength{\tabcolsep}{0.05in}
\tablewidth{-1pt}
\tablecaption{criteria of confidence levels of being physical pairs }
\tablehead{
\colhead{level}&\colhead{UVW differences}  &     \colhead{UVW uncertainties}   &
\colhead{distance differences} & \colhead{age consistency}\\
&(km s$^{-1}$)&(km s$^{-1}$)&(pc)& }
\startdata
`A'& $<$ 15& $<$ 20 & $<$500 & Y \\
`B'& $<$ 15& $<$ 20 & $<$500 & NULL \\
`C'& $<$ 35& $<$ 35 & $<$500 & Y \\
`D'& $<$ 35& $<$ 35 & $<$500 & NULL \\
\enddata
\end{deluxetable}




\end{document}